\documentclass[journal,twoside,web]{ieeecolor}
\usepackage{generic}
\usepackage{cite}
\usepackage{amsmath,amssymb,amsfonts}
\usepackage{algorithm}
\usepackage{graphicx}
\usepackage{hyperref}
\usepackage{multirow}
\hypersetup{hidelinks=true}
\usepackage{textcomp}
\usepackage{booktabs}
\usepackage{array}
\usepackage{tabularx}

\newcolumntype{L}[1]{>{\raggedright\arraybackslash}p{#1}}
\newcommand\etal{\emph{et al.}}
\usepackage[utf8]{inputenc}
\def\BibTeX{{\rm B\kern-.05em{\sc i\kern-.025em b}\kern-.08em
    T\kern-.1667em\lower.7ex\hbox{E}\kern-.125emX}}
\begin{document}
\title{Federated Learning for Large Models in Medical Imaging: A Comprehensive Review}
\author{Mengyu Sun, Ziyuan Yang, Yongqiang Huang, Hui Yu, Yingyu Chen, 
Shuren Qi, \\Andrew Beng Jin Teoh, \IEEEmembership{Senior Member, IEEE}, Yi Zhang, \IEEEmembership{Senior Member, IEEE}
\thanks{This work did not involve human subjects or animals in its research.}
\thanks{\textit{Corresponding author: Yi Zhang}}
\thanks{M. Sun, Z. Yang, Y. Huang, and Y. Zhang are with the School of Cyber Science and Engineering, Sichuan University, Chengdu 610065, China (e-mail: mysun2001999@163.com, cziyuanyang@gmail.com, yqhuang2912@gmail.com, and yzhang@scu.edu.cn).}
\thanks{H. Yu is with Sichuan Institute of Computer Sciences, Chengdu 610042, China (e-mail: smileeudora@163.com).}
\thanks{Y. Chen is with the College of Computer Science, Sichuan University, Chengdu 610065, China (e-mail: cyy262511@gmail.com).}
\thanks{S. Qi is with the Department of Mathematics, The Chinese University of Hong Kong, Hong Kong, China (e-mail: shurenqi@cuhk.edu.hk).}
\thanks{A. B. J. Teoh is with the School of Electrical and Electronic Engineering, College of Engineering, Yonsei University, Seoul, Republic of Korea (e-mail: bjteoh@yonsei.ac.kr).}                                    
}

\maketitle
\begin{abstract}
Artificial intelligence~(AI) has demonstrated considerable potential in the realm of medical imaging. However, the development of high-performance AI models typically necessitates training on large-scale, centralized datasets. This approach is confronted with significant challenges due to strict patient privacy regulations and legal restrictions on data sharing and utilization. These limitations hinder the development of large-scale models in medical domains and impede continuous updates and training with new data. Federated Learning (FL), a privacy-preserving distributed training framework, offers a new solution by enabling collaborative model development across fragmented medical datasets. In this survey, we review FL’s contributions at two stages of the full-stack medical analysis pipeline. First, in upstream tasks such as CT or MRI reconstruction, FL enables joint training of robust reconstruction networks on diverse, multi-institutional datasets, alleviating data scarcity while preserving confidentiality. Second, in downstream clinical tasks like tumor diagnosis and segmentation, FL supports continuous model updating by allowing local fine-tuning on new data without centralizing sensitive images. We comprehensively analyze FL implementations across the medical imaging pipeline, from physics-informed reconstruction networks to diagnostic AI systems, highlighting innovations that improve communication efficiency, align heterogeneous data, and ensure secure parameter aggregation. Meanwhile, this paper provides an outlook on future research directions, aiming to serve as a valuable reference for the field's development.
\end{abstract}


\section{Introduction}
\begin{figure*}
    \centering
    \includegraphics[width=1\linewidth]{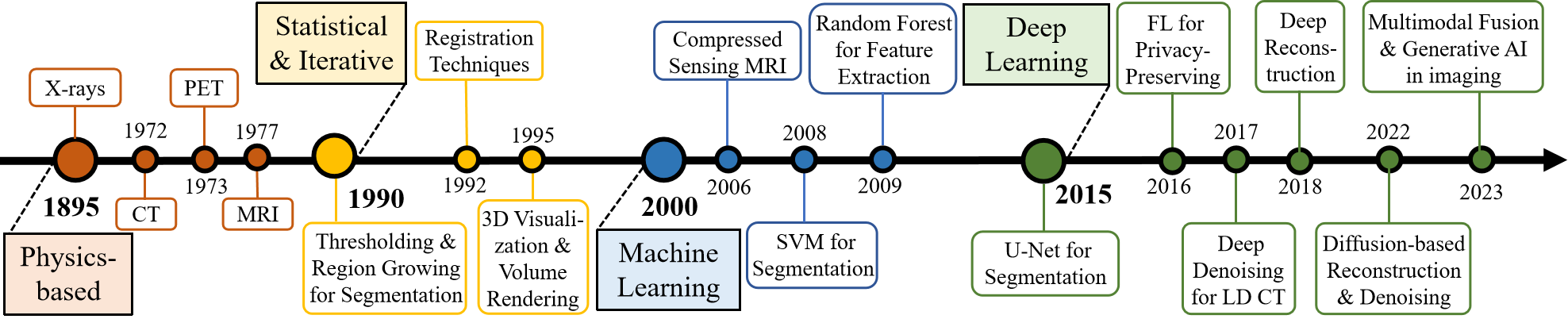}
    \caption{Evolution of Medical Image Analysis and Reconstruction.}
    \label{fig:timeline}
\end{figure*}

In recent years, the rapid advancement of artificial intelligence (AI) has demonstrated immense potential across diverse domains~\cite{xu2021artificial,rashid2024ai,wang2023scientific}. Fig.~\ref{fig:timeline} illustrates the evolution of medical image analysis and reconstruction, highlighting key advancements in the field. In medical imaging, AI technologies are increasingly transforming modern methods for medical image analysis and processing~\cite{pinto2023artificial}. However, most existing AI approaches remain data-driven and require large-scale, high-quality, well-annotated datasets for training~\cite{zha2025data,pasrija2022machine}. Medical image annotation requires manual labeling by domain experts, such as radiologists and pathologists—a process that is both cost-prohibitive and time-intensive~\cite{van2021biological}. 

Unlike medical images, natural images can be labeled by non-experts: crowd workers can handle routine tasks such as object detection, classification, and segmentation. This ease of annotation enables the rapid construction of large-scale datasets via crowdsourcing platforms. In contrast, medical images require well-trained radiologists or clinicians, which inherently limits the speed and scale of labeling. Further complicating the issue, medical images contain sensitive patient information governed by privacy laws, storage limitations, and institutional data governance policies~\cite{kaissis2020secure}. Consequently, the very factors that have fueled the success of large foundation models in general-purpose computer vision, abundant data, and open sharing, are largely absent in the medical domain. Thus, addressing data scarcity while protecting patient privacy remains a critical bottleneck and a pressing avenue for innovation.

\begin{table*}[!t]
\caption{Comparison between recent surveys and this paper}
\centering
\resizebox{1\textwidth}{!}{
\begin{tabular}{llll}
\toprule
\textbf{Paper Title}   & \textbf{Year} & \textbf{Focus/Keywords}    & \textbf{Tasks}                                                  \\ \hline
\begin{tabular}[c]{@{}l@{}}Federated Learning for Rare Disease Detection:\\  A Survey\end{tabular}                                                    & 2023 & \begin{tabular}[c]{@{}l@{}}1. Rare disease detection\\ 2. Data privacy\end{tabular}                                                                         & \begin{tabular}[c]{@{}l@{}}1. Medical image diagnosis\\ 2. Electronic health record (EHR) \\ analytics\end{tabular}                                                   \\ \hline
\begin{tabular}[c]{@{}l@{}}Federated Learning in Medical Image Analysis: \\ A Systematic Survey\end{tabular}                                          & 2023 & \begin{tabular}[c]{@{}l@{}}1. Data heterogeneity\\ 2. Privacy preservation\end{tabular}                                                                     & \begin{tabular}[c]{@{}l@{}}1. Disease diagnosis\\ 2. Multi-site collaboration\\ 3. Benchmarking frameworks\end{tabular}                                               \\ \hline
\begin{tabular}[c]{@{}l@{}}Federated Learning for Medical Image Analysis: \\ A Survey\end{tabular}                                                    & 2024 & \begin{tabular}[c]{@{}l@{}}1. Data heterogeneity\\ 2. Data privacy\\ 3. Client-End and Sever-End learning\end{tabular}                                      & \begin{tabular}[c]{@{}l@{}}1. Medical image analysis\\ 2. Multi-site collaboration\end{tabular}                                                                       \\ \hline
\begin{tabular}[c]{@{}l@{}}Review of Federated Learning and Machine Lear-\\ ning-based Methods for Medical Image Analysis\end{tabular}                & 2024 & \begin{tabular}[c]{@{}l@{}}1. Non-IID data \\ 2. Privacy-preserving\\ 3. FL framework evaluation\\ 4. Cross-modality learning\end{tabular}                  & \begin{tabular}[c]{@{}l@{}}1. Open-source FL framework benchmarking\\ 2. Multi-site FL strategy analysis\\ 3. Medical image analysis taxonomy\end{tabular}            \\ \hline
\begin{tabular}[c]{@{}l@{}}A Survey on Trustworthiness in Foundation \\ Models for Medical Image Analysis\end{tabular}                                & 2024 & \begin{tabular}[c]{@{}l@{}}1. Foundation model\\ 2. Trustworthy AI \\ 3. Medical image analysis\end{tabular}                                                & \begin{tabular}[c]{@{}l@{}}1. Segmentation\\ 2. Medical Q\&A\\ 3. Disease diagnosis\end{tabular}                                                                      \\ \hline
\begin{tabular}[c]{@{}l@{}}Federated Learning in Radiomics: A Comprehen-\\ sive Meta-survey on Medical Image Analysis\end{tabular}                    & 2025 & \begin{tabular}[c]{@{}l@{}}1. Radiomics\\ 2. Meta-survey\\ 3. Medical imaging\\ 4. Privacy preservation\end{tabular}                                        & \begin{tabular}[c]{@{}l@{}}1. Performance analysis of FL models \\ vs. centralized learning approaches\\ 2. Systematic review of FL-MI surveys \\ using PRISMA methodology\end{tabular} \\ \hline
\begin{tabular}[c]{@{}l@{}}A Survey on Federated Learning for Deploying \\ Foundation Medical Models: From Imaging to \\ Diagnosis(Ours)\end{tabular} & -    & \begin{tabular}[c]{@{}l@{}}1. Full-stack clinical analysis\\ 2. Federated Learnig\\ 3. Foundation Model \\ 4. Secure large model deployment\end{tabular} & \begin{tabular}[c]{@{}l@{}}1. Reconstruction\\ 2. Medical image diagnosis\\ 3. Medical image segmentation\end{tabular}                                            \\ \bottomrule
\end{tabular}
}
\vspace{-10pt}
\label{tab:review}
\end{table*}

Collaborative machine learning across multiple data owners, with a focus on preserving data privacy, has garnered substantial attention from both academia and industry. To enable privacy-preserving machine learning, McMahan~\cite{mcmahan2017communication} propose Federated Learning (FL), a distributed learning framework known as FedAvg. Due to its inherent privacy-preserving properties, FL has been widely adopted in various scenarios~\cite{yang2024physics}. In FL, clients independently train local models using their own data and upload model parameters or gradients to a central server. The server aggregates these updates to refine a global model, which is then redistributed to the clients for subsequent training rounds. During the entire training process, clients’ data remains local, with only model parameters or gradient updates transmitted to the central server. This mechanism mitigates data leakage risks and strengthens privacy protection capabilities~\cite{lu2020federated}.

In smart healthcare systems, workflows typically include both upstream medical image reconstruction and downstream image analysis tasks~\cite{hooper2021impact}. However, the nature and impact of data heterogeneity vary significantly between these two categories. 

Medical image reconstruction primarily focuses on restoring high-quality images from low-quality or incomplete imaging data~\cite{gothwal2022computational}. For instance, due to the potential harm of X-ray radiation, clinical protocols often mandate reduced radiation doses during medical examinations. However, this reduction inevitably leads to degraded image quality. 

In low-dose (LD) computed tomography (CT), different healthcare institutions may employ various scanner models or LD protocols (e.g., scanning angles, X-ray photon intensities), resulting in distinct noise distribution patterns~\cite{chen2017low}. This inconsistency in data distribution hinders the generalizability of conventional DL models across clinical settings, consequently affecting reconstruction stability and accuracy. 

In contrast, magnetic resonance imaging (MRI) often adopts accelerated acquisition protocols to shorten scanning time and improve patient comfort. However, variations in MRI scanner hardware configurations (e.g., magnetic field strengths, signal acquisition protocols) and institution-specific reconstruction algorithms collectively contribute to MRI data heterogeneity~\cite{huang2024data}.

Data heterogeneity in medical image analysis primarily stems from three fundamental sources: (1) demographic distribution discrepancies across hospital populations~\cite{aranda2021impact}, (2) variation in histopathological data processing protocols~\cite{de2021residual}, and (3) imbalanced disease prevalence ratios~\cite{yuan2022novel}. Specifically, individual factors such as age, gender, and ethnicity contribute to anatomical variability and diverse lesion characteristics in medical images. Meanwhile, differences in histopathological preparation, including staining protocols and digital scanning devices, further shift data distributions. In addition, disease prevalence varies significantly across institutions: specialized hospitals focus on particular disease groups, while general hospitals serve more diverse populations, with disease severity also varying among medical centers.

Consequently, addressing data heterogeneity~\cite{panayides2020ai} in both imaging and analytical tasks has emerged as a critical research frontier in medical image analysis. The main challenge lies in developing methods that simultaneously mitigate model drift caused by divergent optimization trajectories in FL processes and enhance the generalization of AI models across institutions. To address these challenges, this review provides a technical analysis of state-of-the-art solutions tailored to imaging-oriented and analysis-driven FL frameworks.

In recent years, several comprehensive surveys on FL in medical imaging have been published.
For example, Guan~\etal~\cite{guan2024federated} provide a comprehensive survey of FL methods in medical image analysis, which categorizes approaches into client-side, server-side, and communication techniques. Hernandez-Cruz~\etal~\cite{hernandez2024review} similarly review FL research in medical imaging, highlighting applications (e.g., cardiology, dermatology, oncology) and recurring challenges, such as non-IID data distributions and privacy preservation.
Silva~\etal~\cite{da2023federated} offer a systematic survey on medical imaging modalities (MRI, CT, X‑ray, histology), which discusses the applications, contributions, limitations, and challenges of FL in these domains.
Wang~\etal~\cite{wang2023federated} investigate FL specifically for rare disease detection and summarize existing AI techniques and available datasets for this niche application
In a related area, Shi~\etal~\cite{shi2024survey} explore the trustworthiness aspects of foundation models in medical image analysis, a topic that complements FL surveys by focusing on privacy, robustness, and fairness in large pretrained models. Finally, Raza~\etal~\cite{raza2025federated} conduct a PRISMA-based meta-survey of FL in radiomics, which aggregates trends in tumor detection, organ segmentation, and disease classification tasks.

While these reviews provide valuable insights, their scope often fails to cover the entire imaging pipeline. Existing surveys typically examine image reconstruction, segmentation, and diagnosis as separate topics, rather than treating them as interconnected stages in a federated workflow. Furthermore, the integration of emerging large medical foundation models and advanced data compression techniques has received limited systematic attention in the context of FL across the entire imaging chain. This paper aims to bridge these gaps. We focus on integrating FL throughout the end-to-end medical imaging pipeline, starting from physics-driven image reconstruction to downstream analysis tasks. Additionally, we explore opportunities to incorporate large-scale vision models and efficient data compression techniques into FL frameworks tailored for this comprehensive workflow. A comparison with previous surveys is summarized in Table~\ref{tab:review}, highlighting the expansive scope of this review.

The remainder of this paper is organized as follows:~\ref{chap:Fed} introduces FL workflows and outlines the associated challenges; Chapter~\ref{chap:3} reviews existing FL research in medical image reconstruction; Chapter~\ref{chap:4} analyzes FL applications in medical image analysis; Chapter~\ref{chap:5} elucidates persistent technical bottlenecks and clinical implementation challenges, and proposes future research directions; finally, we summarize the key findings and contributions.

\section{Preliminaries on FL}
\label{chap:Fed}
\begin{figure*}
    \centering
    \includegraphics[width=.8\linewidth]{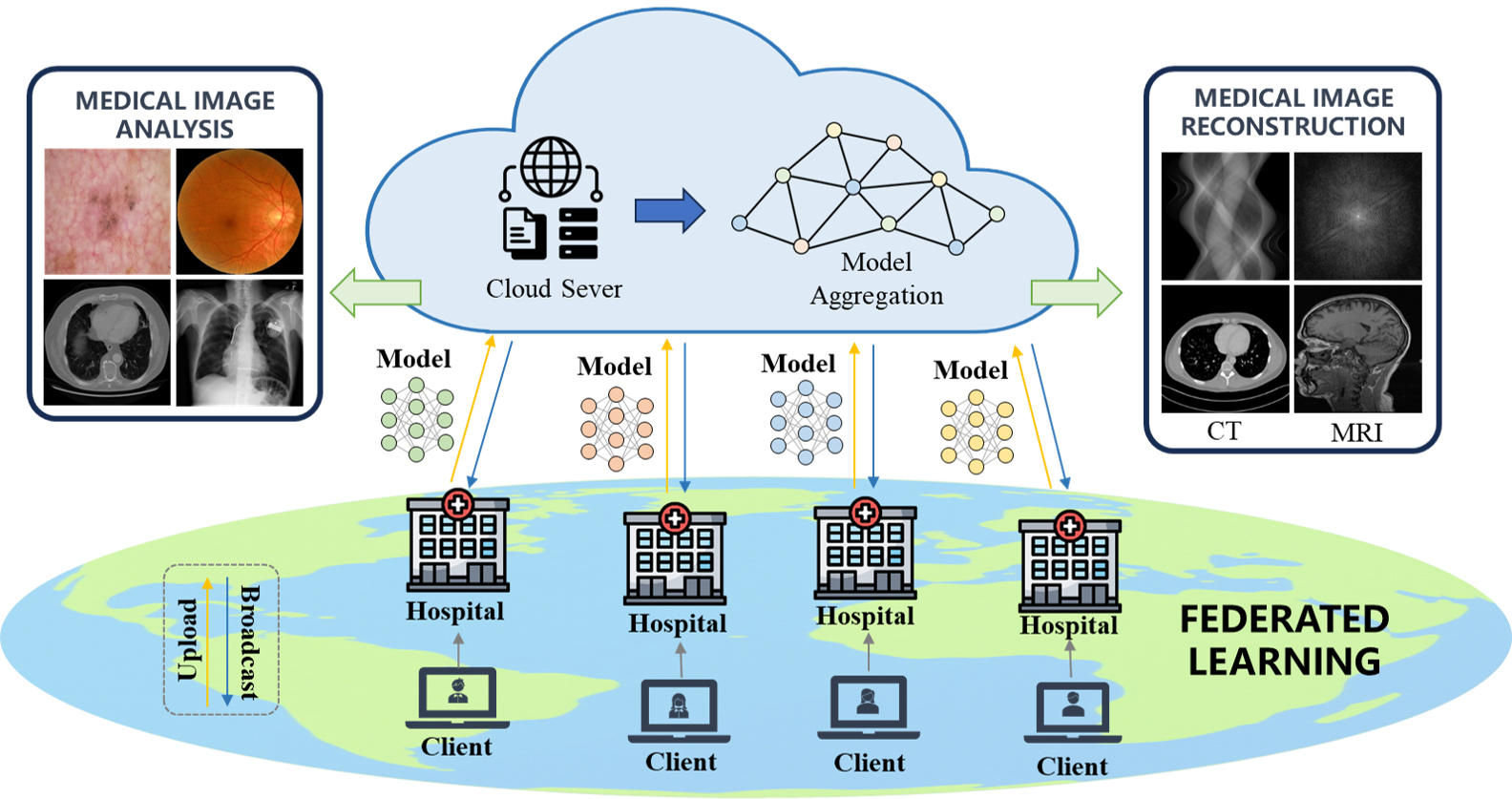}
    \caption{Fical Imaging and Medical Image Analysis.}
    \label{fig1}
\end{figure*}
In medical imaging applications, FL offers a privacy-preserving paradigm that allows decentralized healthcare institutions to jointly train a robust model without sharing patient data. The overall workflow can be organized into four key stages as follows.

\textbf{(1) Initialization}
First, the central server defines the FL task and its requirements, then identifies and invites the participating clients. After establishing collaboration agreements, the server initializes the global model and distributes the initial parameters to all clients, ensuring synchronized initialization of local models throughout the federated network.

\textbf{(2) Local Training}
Upon receiving the model from the server, either during initialization or at the commencement of each communication round, clients leverage local computational resources to train the model on their private datasets, aiming to minimize the loss function of the local model. While some FL approaches may involve multiple local models with distinct loss functions that regularize the divergence between local and global models to mitigate deviation and preserve consistency, this discussion focuses specifically on scenarios that employ a single model architecture. For the sake of analytical uniformity, the loss function considered here includes only task-specific objectives and excludes any additional regularization terms.
The local model update is performed as follows:
\begin{equation} \label{eq:local_update}
\theta^{k} \leftarrow \arg \min _{\theta} \mathcal{L}\left(\theta\right),
\end{equation}
where $\mathcal{L}$ denotes the task-specific loss function (e.g., cross-entropy loss for classification tasks), ${\theta}$ denotes the model parameters, and ${\theta}^{k}$ represents the parameters of the $k$-th client's local model.

\textbf{(3) Model Aggregation}
After receiving local model parameters from multiple clients, the central server aggregates them to generate an updated global model. In the FedAvg algorithm, the global model is updated by performing weighted averaging of parameters uploaded by clients~\cite{collins2022fedavg}. Specifically, FedAvg calculates weighting coefficients based on the data volume of each client, ensuring that clients with larger datasets contribute more proportionally to the global model during aggregation. This mechanism aligns the aggregation process with the relative significance of each client's data distribution to the global model's optimization trajectory.
The following formula represents this process:
\begin{equation} \label{eq:aggregation}
\theta_{t+1}=\sum_{k=1}^{K} \frac{\left|\mathcal{D}_{k}\right|}{\mathcal{|D|}}\theta_{t}^{k},
\end{equation}
where $\theta_{t}^{k}$ denotes the model parameters of the $k$-th client in the $t$-th round, $K$ represents the total number of clients, $|D_{k}|$ and $|D|$ indicate the sample size of the $k$-th client and the total dataset respectively, and $\theta_{t+1}$ represents the global model of the $(t+1)$-th round.

\textbf{(4) Model Update}
After aggregating the parameters, the central server updates the global model and distributes the updated parameters to all clients. Upon receiving the updated global model, clients update their local models and initiate a new round of local training. This iterative process repeats the aggregation-training cycle until the training phase is complete.

The overall FL framework for medical imaging and image analysis is illustrated in Fig.~\ref{fig1}.

Not all FL algorithms strictly follow this standard workflow. Depending on the application context and underlying theoretical assumptions, some methods selectively refine specific components of the pipeline. For instance, knowledge distillation (KD)~\cite{gou2021knowledge} can replace gradient exchange in some cases to leverage heterogeneous client data for training more robust and generalizable models~\cite {zhu2021data}. Despite these variations, all these approaches share a common goal: enabling collaborative learning across distributed data while preserving data privacy.

However, as noted earlier, real-world healthcare settings inherently involve variations in geographic location, population demographics, and clinical protocols across institutions, leading to locally collected data with non-identical distributions~\cite{booth2019real}. As a result, medical data from different organizations often exhibit varying degrees of feature and label shifts. Addressing the challenges posed by non-Independent and Identically Distributed (non-IID) data is crucial for the effective deployment of FL in smart healthcare systems. 

For example, FedProx~\cite{li2020federated} is considered an enhanced version of FedAvg, which introduces a proximal regularizer alongside the task-specific loss during optimization. This regularizer constrains the discrepancy between local and global models to prevent model drift. Similarly, Li~\etal~\cite{li2021model}propose the MOdel-cONtrastive (MOON) method, which minimizes the distance between feature representations of local and global models to mitigate model divergence. The key difference lies in FedProx’s parameter-level constraints versus MOON’s feature-level constraints. Additionally, some approaches mitigate data heterogeneity through optimized optimizer designs~\cite{wang2020tackling}. Beyond parameter transmission in conventional frameworks, some studies employ KD to facilitate global knowledge learning through transmitted knowledge representations~\cite{wu2022communication}.

To address data heterogeneity, several methods have proposed personalized FL. Unlike conventional FL, which trains a single global model, personalized FL allows each client, such as individual hospitals or medical devices, to maintain a model tailored to its local data distribution. For example, Li~\etal~\cite{li2021fedbn} propose FedBN, which alleviates feature shift by retaining client-specific Batch Normalization (BN) layers. While all other model parameters are aggregated globally, BN layers remain localized to preserve institution-specific characteristics.

Shamsian~\etal~\cite{shamsian2021personalized} develop pFedHN, a hypernetwork-based method that generates full local model parameters through a central hypernetwork. Although effective for lightweight models, its parameter count becomes prohibitive in medical imaging tasks, where models tend to be large, leading to performance degradation. Building on the ideas of pFedHN and FedBN, Li~\etal~\cite{li2023fedtp} introduce a method that utilizes hypernetworks to generate personalized projection matrices for self-attention layers, enabling client-specific queries, keys, and values. However, this approach is restricted to Transformer-based architectures, thereby limiting its applicability across diverse model types. To improve collaboration across clients, Zhao~\etal~\cite {zhao2022personalized} propose learning individualized feature spaces, enabling the identification of models compatible with personalized knowledge sharing.

Most of the aforementioned methods are primarily developed for natural image tasks and tend to exhibit suboptimal performance when directly applied to medical imaging. As a result, designing FL algorithms explicitly adapted to the distinctive properties of medical images remains a pressing and unresolved challenge in this field.

\section{Federated Learning in Medical Imaging}
\label{chap:3}
FL in medical imaging, as outlined above, unifies diverse institutions into a privacy‑preserving collaborative network. Unlike generic computer vision problems, medical image reconstruction and analysis are inherently physics-driven and modality-specific. For instance, variations in CT scanner protocols or MRI acquisition parameters directly influence noise patterns and reconstruction stability. Thus, mathematically modeling tasks, such as LDCT denoising or accelerated MRI reconstruction, provides the foundation for developing FL-compatible solutions that reconcile distributed data constraints with clinical fidelity requirements.

\subsection{Problem Modeling}
The medical image reconstruction task~\cite{xi2015simultaneous} can be formulated as:
\begin{equation}
\min_{x} \frac{1}{2} \|Ax - y\|_{2}^{2} + R(x),
\label{eq:3}
\end{equation}
where $\|\cdot\|_{2}^{2}$ denotes the $L_2$ norm, $A \in \mathbb{R}^{m\times n}$ represents the system matrix, $x \in \mathbb{R}^{n}$ is the image to be reconstructed, $y \in \mathbb{R}^{m}$ contains the measurement data, and $R(\cdot)$ denotes the regularization term, typically constructed based on prior knowledge.

Additionally, the LDCT image denoising~\cite{yang2018low} problem can be expressed as:
\begin{equation}
\min_{\omega} \|F(x_l,\omega) - x_n\|_{2}^{2},
\label{eq:4}
\end{equation}
where $x_l$ represents the LDCT image, $x_n$ corresponds to the normal-dose CT image, $F(\cdot,\omega)$ denotes the target model designed to make $x_l$ approximate $x_n$, and $\omega$ represents the parameters of the imaging network.

\subsection{CT Imaging}
\begin{table*}[ht]
\caption{FL Applications for CT Reconstruction}
\centering
\resizebox{\textwidth}{!}{%
\begin{tabular}{lllll}
\toprule
\textbf{Ref.}                                        & \textbf{Target Task}                                                                                         & \textbf{\begin{tabular}[c]{@{}l@{}}Federated \\ Framework\end{tabular}}            & \textbf{Dataset}                                                                        & \textbf{Key Contributions}                                                                                                                                                                                  \\ \hline
\begin{tabular}[c]{@{}l@{}}Yang \textit{et} \\ \textit{al.}~\cite{yang2023hypernetwork}\end{tabular}& \begin{tabular}[c]{@{}l@{}}LDCT Reconstruction\\ and Postprocessing\end{tabular} & HyperFed                                                       & \begin{tabular}[c]{@{}l@{}}IH-AAPM-Mayo Clinic \\ LDCT Dataset\end{tabular}                      & \begin{tabular}[c]{@{}l@{}}Physics-driven hypernetwork modulates global imaging\\ features to mitigate scanner heterogeneity.\end{tabular}                            \\ \hline

\begin{tabular}[c]{@{}l@{}}Yang \textit{et} \\ \textit{al.}~\cite{yang2025patient}\end{tabular} & LDCT Denoising                                                                   & SCAN-PhysFed                                                   & \begin{tabular}[c]{@{}l@{}}NIH-AAPM-Mayo Clinic\\ LDCT Dataset\end{tabular}                      & \begin{tabular}[c]{@{}l@{}}First work to incorporate LLM in FL for CT denoising \\ with dual-level (scanning/anatomy) personalization.\end{tabular}                   \\ \hline

\begin{tabular}[c]{@{}l@{}}Chen \textit{et} \\ \textit{al.}~\cite{chen2023federated}\end{tabular}                                                  & \begin{tabular}[c]{@{}l@{}}Multi-Condition LD-
 \\CT Reconstruction\end{tabular}                & FedCG                                                                          & \begin{tabular}[c]{@{}l@{}}TU-FG-207A-04, Uni-\\ tobrain, kits19, 6 \\ private CT datasets\end{tabular} & \begin{tabular}[c]{@{}l@{}}Federated cross-domain framework for multi-condition\\ CT reconstruction via condition-adaptive sinogram/\\ image processing.\end{tabular} \\ \hline
 
\begin{tabular}[c]{@{}l@{}}Xu \textit{et} \\ \textit{al.}~\cite{xu2025personalized}\end{tabular}                                                     & LDCT Reconstruction                                                                    & \begin{tabular}[c]{@{}l@{}}Personalized \\ Fed-LDCT\end{tabular}               & LoDoPaB, Mayo                                                                                           & \begin{tabular}[c]{@{}l@{}}Personalized FL framework combining client-specific \\ artifact modeling and global reconstruction.\end{tabular}                           \\ \hline

\begin{tabular}[c]{@{}l@{}}Song \textit{et} \\ \textit{al.}~\cite{song2021federated}\end{tabular}                                                & LDCT Denoising                                                                         & \begin{tabular}[c]{@{}l@{}}Federated \\ CycleGAN\end{tabular}                  & AAPM CT dataset                                                                                         & \begin{tabular}[c]{@{}l@{}}Proposes FedCycleGAN with AdaIN for private image \\ translation and lower bandwidth.\end{tabular}                                         \\ \hline

\begin{tabular}[c]{@{}l@{}}Chen \textit{et} \\ \textit{al.}~\cite{chen2024fedfdd}\end{tabular}                                                  & LDCT Denoising                                                                         & FedFDD                                                                         & \begin{tabular}[c]{@{}l@{}}Mayo Clinic (TCIA)\\  Abdomen/Chest/Head\end{tabular}                        & \begin{tabular}[c]{@{}l@{}}Frequency decomposition updates global high-frequency \\ and preserves local low-frequency for non-IID data.\end{tabular}                  \\ \hline

\begin{tabular}[c]{@{}l@{}}Li \textit{et} \\ \textit{al.}~\cite{li2023semi}\end{tabular}                                                   & \begin{tabular}[c]{@{}l@{}}LDCT Image \\ Quality Enhancement\end{tabular}              & \begin{tabular}[c]{@{}l@{}}Semi-centralized\\ FL\end{tabular}                  & \begin{tabular}[c]{@{}l@{}}Multi-institutional\\  CT datasets (3 clients)\end{tabular}                  & \begin{tabular}[c]{@{}l@{}}Introduces SC-FL with server-side centralized dataset \\ training to standardize global model.\end{tabular}                                \\ \hline

\begin{tabular}[c]{@{}l@{}}Li \textit{et} \\ \textit{al.}~\cite{li2023robust}\end{tabular}                                                     & \begin{tabular}[c]{@{}l@{}}Multi-institution \\ LDCT Image \\ Enhancement\end{tabular} & \begin{tabular}[c]{@{}l@{}}Semi-supervised/\\ semi-centralized FL\end{tabular} & \begin{tabular}[c]{@{}l@{}}Multi-institutional \\ LDCT datasets \\ (3 institutions)\end{tabular} & \begin{tabular}[c]{@{}l@{}}Combines semi-supervised FL and centralized fine-tuning \\ to enhance LDCT robustness.\end{tabular}                                        \\
\bottomrule
\end{tabular}%
}
\label{table1}
\end{table*}
The reduction of CT radiation dosage constitutes a critical safety measure to minimize patients' exposure to potential X-ray hazards~\cite{kubo2014radiation}. However, dose reduction inevitably amplifies noise and introduces artifacts that compromise image quality~\cite{seibert2004tradeoffs}. Consequently, recent years have witnessed substantial research efforts dedicated to developing LDCT reconstruction algorithms within FL frameworks, aiming to enhance reconstruction quality while ensuring patient safety.

Contemporary methodologies primarily address the LDCT reconstruction challenge in FL through dual optimization objectives: client-specific adaptation and global imaging feature learning. A representative approach by Yang~\etal~\cite{yang2023hypernetwork}, termed HyperFed, leverages the intrinsic correlation between noise distribution and scanning protocols in LDCT imaging. This method employs a hypernetwork architecture to extract protocol-specific noise characteristics from scanning parameters. To counteract the detrimental effects of data heterogeneity, the framework implements client-specific domain adaptation through personalized hypernetworks that modulate the shared global reconstruction network. This modulation mechanism enables dynamic adaptation to variations in imaging devices and scanning protocols. By explicitly modeling the noise distribution disparities among clients, the approach enhances personalization while improving the generalization capability of FL-optimized LDCT reconstruction systems. 

SCAN-PhysFed~\cite{yang2025patient} introduces dual-level physical modeling that simultaneously addresses scanning protocol variations and patient-specific anatomical features. This framework employs large language models (LLMs) to generate anatomical prompts from radiology reports, which enable anatomy-informed hypernetworks for patient-level adaptation while explicitly modeling physical scanner parameters through orthogonal-constrained hypernetworks.

Chen~\etal~\cite{chen2023federated} develop FedCG to leverage multi-client data through two mechanisms: local client-level sinogram learning and cross-client image reconstruction for conditional generalization. This approach aligns measurement domain features through a conditional generalization network on the server, enabling the learning of latent shared characteristics while preserving client-specific features under different conditions. 

Concurrently, Xu~\etal~\cite{xu2025personalized} propose PerFed-LDCT, a personalized FL framework for CT reconstruction, which employs an artifact fusion network comprising two components: client-specific models for domain-specific artifact extraction and a global shared model for generalized image reconstruction. Complementing these approaches, Song~\etal~\cite{song2021federated} propose Federated CycleGAN for privacy-preserving cross-domain translation, implementing domain-specific loss decomposition and AdaIN-based generators that enable decentralized CT reconstruction across heterogeneous scanner protocols without raw data sharing.

Chen~\etal~\cite{chen2024fedfdd} introduce FedFDD, which uses specialized networks to separately process the high-frequency and low-frequency components of CT data to address spectral domain variations in LDCT scanning protocols. The method improves reconstruction quality by aggregating global high-frequency information while maintaining localized low-frequency characteristics. 

While these methods have demonstrated promising performance, they are generally based on a critical assumption: the availability of perfectly paired LDCT and full-dose CT images~\cite{fletcher2017estimation}. However, in practical clinical scenarios, acquiring precisely matched labeled image pairs remains challenging due to factors such as patient organ motion, variations in scanning times, and equipment-related discrepancies, which consequently restrict the applicability of these approaches.

\subsection{MRI Imaging}
Beyond CT, MRI stands as another crucial imaging modality widely used in clinical practice. The primary objective of rapid MRI imaging is to accelerate scanning speeds while ensuring high image quality~\cite{setsompop2016rapid}. This includes reducing patient scan times, which enhances examination comfort, and leverages advanced signal processing~\cite{lin2024designing} and DL techniques to reconstruct undersampled data, suppress artifacts, and improve resolution and contrast~\cite{lee2024deep}. Recent advancements in federated MRI reconstruction have focused on addressing domain shifts, multi-modal harmonization, computational efficiency, and privacy preservation through integrated technical strategies.

A key challenge of MRI imaging lies in mitigating domain discrepancies caused by heterogeneous scanners and imaging protocols. Researchers have proposed several innovative strategies to address domain discrepancies through adaptive personalization. For example, Lyu~\etal~\cite{lyu2023adaptive} introduce ACM-FedMRI, a dual mechanism that leverages client-specific hypernetworks to generate adaptive channel-selection weights from client ID embeddings, thereby enabling personalized feature recalibration and preserving high-frequency recovery. 

Similarly, Feng~\etal~\cite{feng2022specificity} decouple the reconstruction model into a globally shared encoder and client-specific decoders, facilitating personalized collaborative reconstruction and adaptive parameterization of decoding layers to mitigate domain shifts. In addition, Guo~\etal~\cite{guo2021multi} propose the FL-MRCM framework, which uses adversarial training between local reconstruction networks and a shared domain identifier to align intermediate feature distributions across source and target sites, thus achieving domain-invariant reconstruction.

Federated frameworks have also focused on multi-modal and cross-protocol harmonization by leveraging cross-modal knowledge transfer and modality synthesis. Yan~\etal~\cite{yan2023federated} develop Fed-PMG, which addresses modality deficiency by clustering amplitude spectra from multi-modal participants into centroids to enable missing modality synthesis through local phase retention and centroid-based amplitude interpolation. 

In parallel, pFLSynth~\cite{dalmaz2024one} tackles intra-modality heterogeneity in multi-contrast MRI synthesis by designing personalized normalization and attention blocks to adaptively modulate feature statistics for individual sites and contrast translation tasks, while partially aggregating later generator stages to balance generalization and specialization. 

These two frameworks demonstrate divergent strategies: Fed-PMG emphasizes cross-modal consistency through spectral alignment, whereas pFLSynth prioritizes site- and task-specific adaptation within a unified model architecture, both aiming to enhance robustness against heterogeneous data distributions in federated medical imaging. Vertical FL frameworks, such as Fed-CRFD~\cite{yan2024cross}, also contribute by disentangling modality-invariant and modality-specific features through adversarial training, thereby reducing domain shifts caused by heterogeneous imaging protocols while preserving data privacy.

\begin{table*}[ht]
\caption{FL Applications for MRI Reconstruction}
\centering
\resizebox{\textwidth}{!}{%
\begin{tabular}{lllll}
\toprule
\textbf{Ref.}                                        & \textbf{Target Task}                                                                                         & \textbf{\begin{tabular}[c]{@{}l@{}}Federated \\ Framework\end{tabular}}            & \textbf{Dataset}                                                                        & \textbf{Key Contributions}                                                                                                                                                                        \\ \hline
\begin{tabular}[c]{@{}l@{}}Lyu \textit{et} \\ \textit{al.}~\cite{lyu2023adaptive}\end{tabular} & MRI Reconstruction                                                                                           & ACM-FedMRI                                                                         & \begin{tabular}[c]{@{}l@{}}BraTS, fastMRI, IXI,\\ Volunteer\end{tabular}                & \begin{tabular}[c]{@{}l@{}}Client-specific hypernetworks enable adaptive MRI \\ reconstruction via dynamic channel decoupling.\end{tabular}                          \\ \hline

\begin{tabular}[c]{@{}l@{}}Feng \textit{et} \\ \textit{al.}~\cite{feng2022specificity}\end{tabular}                                             & MRI Reconstruction                                                                                           & FedMRI                                                                             & \begin{tabular}[c]{@{}l@{}}fastMRI, BraTS, SMS,\\  uMR\end{tabular}                     & \begin{tabular}[c]{@{}l@{}}Preserves client-specific features while mitigating\\ domain shifts in federated MR reconstruction.\end{tabular}                                                       \\ \hline

\begin{tabular}[c]{@{}l@{}}Guo \textit{et} \\ \textit{al.}~\cite{guo2021multi}\end{tabular}                                                 & MRI Reconstruction                                                                                           & \begin{tabular}[c]{@{}l@{}}FedAvg with Cross-\\ site Modeling\end{tabular}         & \begin{tabular}[c]{@{}l@{}}fastMRI, HPKS, IXI, \\ BraTS\end{tabular}                    & \begin{tabular}[c]{@{}l@{}}FL-MRCM aligns latent features across institutions\\ to address domain shift.\end{tabular}                                                                             \\ \hline

\begin{tabular}[c]{@{}l@{}}Yan \textit{et} \\ \textit{al.}~\cite{yan2023federated}\end{tabular}                                                  & \begin{tabular}[c]{@{}l@{}}Multi-modal MRI\\ Reconstruction under \\ Modality Missing\end{tabular}           & Fed-PMG                                                                            & fastMRI, uMR                                                                            & \begin{tabular}[c]{@{}l@{}}Cluster-enhanced modality completion with \\ communication efficiency.\end{tabular}                                                                                    \\ \hline

\begin{tabular}[c]{@{}l@{}}Dalmaz \textit{et} \\ \textit{al.}~\cite{dalmaz2024one}\end{tabular}                                               & \begin{tabular}[c]{@{}l@{}}Multi-contrast MRI\\ Synthesis\end{tabular}                                       & pFLSynth                                                                           & \begin{tabular}[c]{@{}l@{}}IXI, BRATS, MIDAS, \\ OASIS\end{tabular}                     & \begin{tabular}[c]{@{}l@{}}Introduces personalized FL with site/task-specific \\ modulation blocks and partial aggregation to \\ address data heterogeneity.\end{tabular}                         \\ \hline

\begin{tabular}[c]{@{}l@{}}Yan \textit{et} \\ \textit{al.}~\cite{yan2024cross}\end{tabular}                                                  & \begin{tabular}[c]{@{}l@{}}Cross-modal MRI\\ Reconstruction\end{tabular}                                     & Federated-CRFD                                                                     & \begin{tabular}[c]{@{}l@{}}fastMRI, Private\\ clinical dataset\end{tabular}             & \begin{tabular}[c]{@{}l@{}}Proposes vertical FL framework with feature disen-\\ tanglement and cross-client latent alignment for \\ multi-modal MRI reconstruction.\end{tabular}                  \\ \hline

\begin{tabular}[c]{@{}l@{}}Feng \textit{et} \\ \textit{al.}~\cite{feng2023learning}\end{tabular}                                                & MRI Reconstruction                                                                                           & FedPR                                                                              & \begin{tabular}[c]{@{}l@{}}fastMRI, FeTS, IXI, \\ Clinical Datasets\end{tabular}        & \begin{tabular}[c]{@{}l@{}}Proposes FedPR to learn visual prompts in null space \\ of global prompts, reducing communication costs and \\ catastrophic forgetting with limited data.\end{tabular} \\ \hline

\begin{tabular}[c]{@{}l@{}}Wu \textit{et} \\ \textit{al.}~\cite{wu2024generalizable}\end{tabular}                                                   & \begin{tabular}[c]{@{}l@{}}Accelerated MRI\\ Reconstruction\end{tabular}                                     & GAutoMRI                                                                           & \begin{tabular}[c]{@{}l@{}}fastMRI, cc359, MoDL-\\ Brain, \\ In-house\end{tabular}      & \begin{tabular}[c]{@{}l@{}}Combines neural architecture search with fairness \\ adjustment for improved generalization and light-\\ weight models in FL-based MRI reconstruction.\end{tabular}    \\ \hline

\begin{tabular}[c]{@{}l@{}}Elmas \textit{et} \\ \textit{al.}~\cite{elmas2022federated}\end{tabular}                                                & \begin{tabular}[c]{@{}l@{}}Accelerated MRI\\ Reconstruction\end{tabular}                                     & FedGIMP                                                                            & \begin{tabular}[c]{@{}l@{}}IXI, fastMRI, BRATS,\\ In-House\end{tabular}                 & \begin{tabular}[c]{@{}l@{}}Proposes FL of generative image priors decoupled \\ from imaging operators to improve cross-site \\ generalization.\end{tabular}                                       \\ \hline

\begin{tabular}[c]{@{}l@{}}Wu \textit{et} \\ \textit{al.}~\cite{wu2024model}\end{tabular}                                                   & \begin{tabular}[c]{@{}l@{}}Accelerated MRI Recons-\\ truction from Under\\ sampled K-space Data\end{tabular} & Model-based FL                                                                     & \begin{tabular}[c]{@{}l@{}}In-house dataset, \\ fastMRI, CC359\end{tabular}             & \begin{tabular}[c]{@{}l@{}}Proposes a model-driven FL framework with adaptive \\ aggregation and attention mechanisms for improved \\ MRI reconstruction under data heterogeneity.\end{tabular}   \\ \hline

\begin{tabular}[c]{@{}l@{}}Ahmed \textit{et} \\ \textit{al.}~\cite{ahmed2025fame}\end{tabular}                                                & \begin{tabular}[c]{@{}l@{}}MRI Reconstruction \\ from Undersampled \\ K-space Data\end{tabular}              & \begin{tabular}[c]{@{}l@{}}FAME with adaptive \\ aggregation strategy\end{tabular} & \begin{tabular}[c]{@{}l@{}}fastMRI Brain, fastMRI \\ Knee, BraTS 2020, IXI\end{tabular} & \begin{tabular}[c]{@{}l@{}}Proposes FL-GAN framework with dynamic weighting \\ and privacy-preserving mechanisms for MRI \\ reconstruction.\end{tabular}                                          \\ \hline

\begin{tabular}[c]{@{}l@{}}Zou \textit{et} \\ \textit{al.}~\cite{zou2023self}\end{tabular}                                                  & \begin{tabular}[c]{@{}l@{}}Accelerated MRI\\ Reconstruction\end{tabular}                                     & \begin{tabular}[c]{@{}l@{}}Self-Supervised FL \\ with centralized learning\end{tabular}              & \begin{tabular}[c]{@{}l@{}}FastMRI, CC359, Modl, \\ In-house\end{tabular}               & \begin{tabular}[c]{@{}l@{}}Enables FL without fully-sampled data using physics-\\ based contrastive networks and soft model updates to \\ handle data heterogeneity.\end{tabular}                 \\ \hline

\begin{tabular}[c]{@{}l@{}}Ahmed \textit{et} \\ \textit{al.}~\cite{ahmed2025fedgraphmri}\end{tabular}                                                & \begin{tabular}[c]{@{}l@{}}MRI Reconstruction\\ under Non-IID Data\end{tabular}                              & GNN, Personalized FL                                                               & IXI, fastMRI                                                                            & \begin{tabular}[c]{@{}l@{}}Developes Louvain-based subgraph partitioning and \\ parameter decomposition for federated MRI recons-\\ truction.\end{tabular}                                                                    \\\bottomrule

\end{tabular}%
}
\label{table2}
\end{table*}

In response to the practical constraints of FL deployments in clinical environments, recent works have prioritized communication efficiency and lightweight architectures. FedPR~\cite{feng2023learning} utilizes prompt-based learning to communicate only compact visual prompts while freezing backbone networks, significantly reducing transmission costs. Meanwhile, GAutoMRI~\cite{wu2024generalizable} automates neural architecture search for physics-informed models using parameter-efficient dilated convolution.
FedGIMP~\cite{elmas2022federated} further decouples global generative priors from site-specific acquisition physics and enables collaborative training without the need to share sensitive coil sensitivity maps.

Another key trend is the incorporation of MRI physics into federated frameworks to enhance the fidelity of reconstruction. ModFed~\cite{wu2024model} integrates unfolded neural networks with MR physics priors. It uses adaptive dynamic aggregation and spatial Laplacian attention to bolster edge recovery and reconstruction performance. Elmas~\etal~\cite{elmas2022federated} propose a two-stage approach in FedGIMP, where cross-client learning is first used to generate global MRI priors through adversarial models, which are then incorporated into the imaging network for personalized reconstruction. 

Privacy preservation remains a critical concern in FL. Ahmed~\etal~\cite{ahmed2025fame} integrate differential privacy (DP)~\cite{dwork2006differential} with encrypted aggregation and use GAN-based generators to decouple raw data from the collaborative training process. Other methods, such as SSFedMRI~\cite{zou2023self}, combine self-supervised contrastive learning~\cite{tian2020makes} with lightweight MoDL (Model-based deep learning) architectures. They also use physics-based supervisory signals generated from $k$-space re-undersampling, which help reduce the reliance on fully sampled data. FedGraphMRI-Net~\cite{ahmed2025fedgraphmri} demonstrates a novel application of graph neural networks by partitioning MRI data into spatially coherent subgraphs through Louvain clustering, which minimizes raw data exposure while enhancing the modeling of anatomical correlations.

Levac~\etal~\cite{levac2023federated} further tackle data scarcity in low-data regimes by employing adaptive optimization algorithms, such as Scaffold~\cite{karimireddy2020scaffold} and FedAdam~\cite{reddi2020adaptive}, to mitigate client drift in non-IID settings through momentum-based updates and dynamic learning rate tuning. As a result, they achieve structural similarity index metrics comparable to centralized training despite limited communication rounds.

In summary, FL introduces a collaborative paradigm for MRI reconstruction by enabling the privacy-preserving integration of multi-institutional data. By incorporating adaptive domain alignment, lightweight network architectures, and physics-informed priors, FL facilitates robust and generalizable MRI reconstruction across diverse and heterogeneous clinical settings.

\section{Federated Learning in Medical Image Analysis}
\label{chap:4}
In smart healthcare, beyond the upstream medical imaging tasks discussed earlier, downstream medical image analysis represents another vital facet of FL in medical imaging. Based on specific clinical objectives, these analysis tasks can be broadly classified into two categories: medical image diagnosis and medical image segmentation. The following sections provide a detailed overview of several representative applications in each category. Related works are summarized in Tables~\ref{table3} and ~\ref{table4}.

\subsection{Disease Detection and Diagnosis}
\begin{table*}[]
\caption{FL Applications for Disease Detection and Diagnosis in Medical Image}
\centering
\resizebox{\textwidth}{!}{%
\begin{tabular}{lllll}
\toprule
\textbf{Ref.} & \textbf{Target Task}                                                                                          & \textbf{\begin{tabular}[c]{@{}l@{}}Federated \\ Framework\end{tabular}}                                        & \textbf{Dataset}                                                                                                   & \textbf{Key Contributions}    \\\hline

\begin{tabular}[c]{@{}l@{}}Palash \textit{et} \\ \textit{al.}~\cite{palash2024federated}\end{tabular}               & \begin{tabular}[c]{@{}l@{}}Lung Tumor \\ Detection\end{tabular}                                            & \begin{tabular}[c]{@{}l@{}}Federated Semi-\\ Supervised Learning\\ with Dynamic\\ Update Strategy\end{tabular} & \begin{tabular}[c]{@{}l@{}}GDPH, TJCH, CHSUMC,\\ RIDER, INTEROBS, LUNG1\end{tabular}                               & \begin{tabular}[c]{@{}l@{}}Dynamic model aggregation leveraging data\\ quality/quantity improves multi-center lung \\ tumor detection.\end{tabular}                                                \\ \hline

 \begin{tabular}[c]{@{}l@{}}Kumar \textit{et} \\ \textit{al.}~\cite{kumar2021blockchain}\end{tabular}                & \begin{tabular}[c]{@{}l@{}}COVID-19 Detection\\ from CT\end{tabular}                                          & \begin{tabular}[c]{@{}l@{}}Blockchain-based\\ FL with capsule\\ networks\end{tabular}                          & CC-19                                                                                                              & \begin{tabular}[c]{@{}l@{}}Combines blockchain FL with spatial normalization\\ and capsule networks for privacy-preserving COVID\\ -19 diagnosis across heterogeneous CT scanners.\end{tabular}       \\ \hline

 \begin{tabular}[c]{@{}l@{}}Lai \textit{et} \\ \textit{al.}~\cite{lai2022federated}\end{tabular}               & \begin{tabular}[c]{@{}l@{}}Automated Detection\\ of COVID-19 Lung\\ Abnormalities in CT\end{tabular}          & FedAvg                                                                                                         & \begin{tabular}[c]{@{}l@{}}Different hospitals'\\ data\end{tabular}                                                & \begin{tabular}[c]{@{}l@{}}Demonstrates privacy-preserving FL framework\\ achieving cross-national generalizability in \\ COVID-19 CT diagnosis.\end{tabular}                                         \\ \hline
 
  \begin{tabular}[c]{@{}l@{}}Heidari \textit{et} \\ \textit{al.}~\cite{heidari2023new}\end{tabular}              & Lung Cancer Detection                                                                                         & \begin{tabular}[c]{@{}l@{}}Blockchain-\\ enabled FL\end{tabular}                                               & \begin{tabular}[c]{@{}l@{}}CIA, KDSB, LUNA16, \\ Local dataset\end{tabular}                                  & \begin{tabular}[c]{@{}l@{}}Integrates blockchain with FL and CapsNets for\\ privacy-preserving lung cancer detection using \\ CT scans.\end{tabular}                                                  \\ \hline
  
 \begin{tabular}[c]{@{}l@{}}Li \textit{et} \\ \textit{al.}~\cite{li2024sift}\end{tabular}               & \begin{tabular}[c]{@{}l@{}}Medical Image\\ Classification\end{tabular}                                        & Serial Framework                                                                                               & \begin{tabular}[c]{@{}l@{}}HAM10000, OrganCMNIST,\\ OrganSMNIST\end{tabular}                                       & \begin{tabular}[c]{@{}l@{}}Proposes a serial FL framework with continual  \\learning and  biomedical language model\\guidance to reduce communication costs.\end{tabular}                                             \\ \hline
 
 \begin{tabular}[c]{@{}l@{}}Zhou \textit{et} \\ \textit{al.}~\cite{zhou2024distributed}\end{tabular}               & \begin{tabular}[c]{@{}l@{}}MRI Brain\\ Tumor Detection\end{tabular}                                           & \begin{tabular}[c]{@{}l@{}}FedAvg, \\ ResNet-50\end{tabular}                                                   & \begin{tabular}[c]{@{}l@{}}Kaggle Brain Tumor\\  Classification (MRI)\end{tabular}                                 & \begin{tabular}[c]{@{}l@{}}Demonstrates FL's effectiveness in privacy-\\ preserving MRI tumor detection using Efficient-\\ Net-B0 on heterogeneous data.\end{tabular}                                 \\ \hline
 
 \begin{tabular}[c]{@{}l@{}}Li \textit{et} \\ \textit{al.}~\cite{li2020multi}\end{tabular}               & \begin{tabular}[c]{@{}l@{}}Medical Image\\ Classification\end{tabular}                                        & \begin{tabular}[c]{@{}l@{}}MLP, MoE\\ modules\end{tabular}                                                     & ABIDE I                                                                                                            & \begin{tabular}[c]{@{}l@{}}Integrates privacy-preserving FL with cross-site \\ domain adaptation to improve ASD classification \\ and detect neuroimaging biomarkers.\end{tabular}                    \\ \hline
 
 \begin{tabular}[c]{@{}l@{}}Bercea \textit{et} \\ \textit{al.}~\cite{bercea2022federated}\end{tabular}               & \begin{tabular}[c]{@{}l@{}}Brain MRI\\ Anomaly Detection\end{tabular}                                         & \begin{tabular}[c]{@{}l@{}}Disentangled \\ global-local \\ parameter learning\end{tabular}                     & \begin{tabular}[c]{@{}l@{}}OASIS, ADNI-S, ADNI-P,\\ KRI, MSLUB, MSISBI,\\ MSKRI, WMH, GBKRI, \\ BRATS\end{tabular} & \begin{tabular}[c]{@{}l@{}}Proposes federated disentangled representation \\ learning to improve cross-institutional anomaly \\ detection without sharing private data or\\ annotations.\end{tabular} \\ \hline
 
 \begin{tabular}[c]{@{}l@{}}Alhamoud  \\ \textit{et al.}~\cite{alhamoud2024fedmedicl}\end{tabular}               & \begin{tabular}[c]{@{}l@{}}Medical Image Classi-\\ fication under Multiple\\ Distribution Shifts\end{tabular} & \begin{tabular}[c]{@{}l@{}}FedAvg, \\ ResNet-18\end{tabular}                                                   & \begin{tabular}[c]{@{}l@{}}CheXpert, Fitzpatrick17k,\\ HAM10000, OL3I, PAPILA,\\ CheXCOVID\end{tabular}            & \begin{tabular}[c]{@{}l@{}}Introduces unified benchmark for evaluating FL\\ under simultaneous label, demographic, and\\  temporal shifts in medical imaging.\end{tabular}                            \\ \hline
 
 \begin{tabular}[c]{@{}l@{}}Kaissis \textit{et} \\ \textit{al.}~\cite{kaissis2021end}\end{tabular}               & \begin{tabular}[c]{@{}l@{}}Pediatric Chest\\ X-ray Classification\end{tabular}                                & PriMIA                                                                                                         & \begin{tabular}[c]{@{}l@{}}Public pediatric\\ pneumonia dataset\end{tabular}                                       & \begin{tabular}[c]{@{}l@{}}Proposes the framework enabling end-to-end \\ encrypted inference for medical imaging FL.\end{tabular}                                                                     \\ \hline
 
 \begin{tabular}[c]{@{}l@{}}Dayan \textit{et} \\ \textit{al.}~\cite{dayan2021federated}\end{tabular}               & \begin{tabular}[c]{@{}l@{}}Predict 24h/72h\\ Oxygen Requirements\\ in COVID-19 Patients\end{tabular}          & Client-server                                                                                                  & \begin{tabular}[c]{@{}l@{}}EXAM-COVID\\ (Chest X-ray)\end{tabular}                                                 & \begin{tabular}[c]{@{}l@{}}Proposes EXAM framework for cross-site clinical\\ outcome prediction.\end{tabular}                                                                                          \\ \hline
 
 \begin{tabular}[c]{@{}l@{}}Xie \textit{et} \\ \textit{al.}~\cite{xie2024mh}\end{tabular}               & \begin{tabular}[c]{@{}l@{}}Breast Cancer \\ \& OCT Disease\\ Classification\end{tabular}                      & MH-pFLGB                                                                                                       & \begin{tabular}[c]{@{}l@{}}BreaKHis, OCT2017, \\ ColonDB, ETIS, ClinicDB,\\  Kvasir-SEG\end{tabular}               & \begin{tabular}[c]{@{}l@{}}Proposes global bypass strategy with feature\\ fusion to handle model heterogeneity without \\ public datasets.\end{tabular}                                     \\\bottomrule
 
\end{tabular}%
}
\label{table3}
\end{table*}

FL frameworks have shown notable effectiveness in privacy-preserving disease diagnosis across distributed CT and MRI datasets~\cite{ali2022federated, qayyum2022collaborative}, which successfully addresses key challenges such as data heterogeneity and limited annotations. Early approaches adopt hybrid architectures that integrated transfer learning with FL to mitigate cross-institutional feature variability. For example, in CT-based lung cancer detection, Palash~\etal~\cite{palash2024federated} propose a dual-phase framework in which transfer learning is first applied to identify optimal feature extractors, such as MobileNetV2~\cite{sandler2018mobilenetv2}, using centralized data. Subsequently, FL is employed with institution-specific preprocessing techniques, including adaptive resizing and augmentation through flipping and rotation, to align with local workflows. This decoupled strategy separates initial feature learning from federated optimization, which enables effective handling of class imbalance in rare cancer subtypes via weighted model aggregation.

For COVID-19 detection, several federated frameworks integrate neural architectures to address cross-site heterogeneity. Kumar~\etal~\cite{kumar2021blockchain} introduce a blockchain-FL hybrid model that employs 3D SegCaps~\cite{lalonde2018capsules} for lesion segmentation and Capsule Networks~\cite{sabour2017dynamic} for classification, capturing spatial hierarchies of ground-glass opacities via dynamic routing. Scanner-induced variability is mitigated through spatial resampling and lung window standardization, while blockchain consensus mechanisms ensure tamper-proof gradient verification using cryptographic hashing.

In parallel, Lai~\etal~\cite{lai2022federated} propose a communication-efficient approach that replaces conventional weight transmission with global feature vector averaging. By incorporating contrastive learning, the method aligns client-specific CT features with class prototypes, which enhances inter-client consistency. The framework supports heterogeneous local models, including ResNet and lightweight CNNs, and addresses computational disparities across institutions by jointly optimizing cross-entropy and contrastive feature losses.

For lung cancer detection, FL frameworks integrate transfer learning for initial feature extraction with decentralized optimization for collaborative training. Blockchain-enhanced FL systems~\cite{heidari2023new} enhance security by encrypting parameter exchanges and employ adaptive histogram equalization to standardize CT images acquired through diverse scanning protocols. This decoupled strategy isolates feature initialization from federated refinement and enables effective handling of class imbalance in rare cancer subtypes through weighted aggregation.

Beyond disease-specific applications, serial FL paradigms such as SiFT~\cite{li2024sift} incorporate continual learning to support multi-class classification. SiFT aligns text-guided feature projections with biomedical semantics, which addresses class imbalance in CT analysis by reinforcing semantic consistency. In brain tumor diagnosis, distributed frameworks~\cite{zhou2024distributed} adopt EfficientNet architectures with dynamic regularization to mitigate non-IID data effects. These frameworks apply adaptive aggregation and preprocessing techniques, such as random cropping, to enhance generalization across heterogeneous clinical sites.

Federated approaches have also been extended to cross-modality tasks, which facilitates adaptation to heterogeneous data sources through techniques such as modality-specific preprocessing and domain-aware aggregation. In fMRI-based autism spectrum disorder classification, mixture-of-experts (MoE) architectures~\cite{masoudnia2014mixture} combine with adversarial domain alignment dynamically adjust to scanner-specific features while ensuring privacy via gradient randomization~\cite{li2020multi}.

In brain MRI anomaly detection, parameter disentanglement strategies separate global anatomical representations from client-specific intensity variations, while latent contrastive learning enforces consistency across intensity-augmented scans by aligning shape-based embeddings. Simultaneously, self-supervised inpainting reduces false positives caused by acquisition artifacts and allows collaborative learning of anatomical patterns without exposing raw data~\cite{bercea2022federated}.

Holistic frameworks, such as FedMedICL~\cite{alhamoud2024fedmedicl}, further address multi-source distribution shifts, including label imbalance and demographic variability, by incorporating class-balancing techniques and adaptive batch normalization. These strategies enhance model robustness in dynamic clinical contexts, including emerging disease scenarios like COVID-19.

For chest X‑ray scans, Kaissis~\etal~\cite{kaissis2021end} propose PriMIA, an open‑source platform that combines secure aggregation, DP gradient descent, and secure multi‑party computation to enable encrypted training and remote inference on pediatric chest radiographs, achieving expert‑level classification performance without exposing raw imaging data. Similarly, Dayan~\etal~\cite{dayan2021federated} develop EXAM, a client–server FL model that jointly leverages chest X‑ray and electronic medical record features from 20 international centers to predict oxygen requirements in COVID‑19 patients, demonstrating significant improvements in both accuracy and cross‑site generalizability compared to locally trained models~\cite{dayan2021federated}. These end‑to‑end privacy‑preserving systems underscore the potential of FL to deliver robust, secure AI‑driven diagnostics and prognostics in real‑world, multi‑institutional healthcare networks.

For classification and segmentation tasks across heterogeneous medical imaging modalities, Xie~\etal~\cite{xie2024mh} propose MH‑pFLGB, a model‑heterogeneous personalized federated learning framework that combines a lightweight global bypass model with a feature‑weighted fusion module to reconcile both statistical and system heterogeneity among clients with diverse network architectures, improving classification and segmentation performance without relying on public datasets.

\subsection{Medical Image Segmentation}

\begin{table*}[t]
\caption{FL Applications for Medical Image Segmentation}
\centering
\renewcommand{\arraystretch}{1.1}
\begin{tabular}{lllll}
\toprule
\textbf{Ref.} & \textbf{Target Task}                                                                                                        & \textbf{\begin{tabular}[c]{@{}l@{}}Federated \\ Framework\end{tabular}}                                          & \textbf{Dataset}                                                                                                            & \textbf{Key Contributions}                                                                                                                                                                                          \\ \hline
\begin{tabular}[c]{@{}l@{}}Kim \textit{et} \\ \textit{al.}~\cite{kim2024federated}\end{tabular}              & \begin{tabular}[c]{@{}l@{}}Multi-organ\\ Segmentation\end{tabular}                                                          & \begin{tabular}[c]{@{}l@{}}FedAvg with global/\\ local KD\end{tabular}                                           & LiTS, KiTS19, MSD, BTCV                                                                                                     & \begin{tabular}[c]{@{}l@{}}Combines global model distillation and local\\ expert distillation to address catastrophic\\ forgetting in partially labeled FL.\end{tabular}                                            \\ \hline

\begin{tabular}[c]{@{}l@{}}Tolle \textit{et} \\ \textit{al.}~\cite{tolle2024funavg}\end{tabular}              & \begin{tabular}[c]{@{}l@{}}Multi-organ\\ Segmentation\end{tabular}                                                          & \begin{tabular}[c]{@{}l@{}}FedAvg, Uncertainty-\\ weighted Head Ensemble\end{tabular}                            & \begin{tabular}[c]{@{}l@{}}LiTS, AMOS, BCV,\\ AbdomenCT-1K, CHAOS,\\ TotalSegmentator,\\ Learn2Reg\end{tabular}             & \begin{tabular}[c]{@{}l@{}}Leverages predictive uncertainty to aggregate\\ segmentation heads for unannotated structures\\ through FL.\end{tabular}                                                                 \\ \hline

\begin{tabular}[c]{@{}l@{}}Wang \textit{et} \\ \textit{al.}~\cite{wang2024feddus}\end{tabular}               & \begin{tabular}[c]{@{}l@{}}Lung Tumor Seg-\\ mentation (CT)\end{tabular}                                                    & \begin{tabular}[c]{@{}l@{}}Federated Semi-\\ Supervised Learning \\ with Dynamic Update \\ Strategy\end{tabular} & \begin{tabular}[c]{@{}l@{}}GDPH, TJCH, CHSUMC,\\ RIDER, INTEROBS, LUNG1\end{tabular}                                        & \begin{tabular}[c]{@{}l@{}}Dynamic model aggregation leveraging data\\ quality/quantity improves multi-center lung\\ tumor segmentation.\end{tabular}                                                               \\ \hline

\begin{tabular}[c]{@{}l@{}}Luo \textit{et} \\ \textit{al.}~\cite{luo2023influence}\end{tabular}              & \begin{tabular}[c]{@{}l@{}}Liver Tumor Seg-\\ mentation (CT) and\\ Brain Tumor Seg-\\ mentation (MRI)\end{tabular}          & \begin{tabular}[c]{@{}l@{}}Server-client\\ architecture with\\ Fed-Avg algorithm\end{tabular}                    & FILTS, FeTS                                                                                                                 & \begin{tabular}[c]{@{}l@{}}Demonstrates strong negative correlation\\ between data distribution distances and Fed-DL\\ performance in tumor segmentation.\end{tabular}                                              \\ \hline

\begin{tabular}[c]{@{}l@{}}Hsiao \textit{et} \\ \textit{al.}~\cite{hsiao2024precision}\end{tabular}              & \begin{tabular}[c]{@{}l@{}}Liver and Tumor \\ Segmentation, HCC \\ detection\end{tabular}                                   & FedAvg                                                                                                           & LiTS Challenge                                                                                                              & \begin{tabular}[c]{@{}l@{}}Proposes Hybrid-ResUNet combining 2D/3D \\models with FL framework to achieve privacy-\\preserving segmentation.\end{tabular}                                                            \\ \hline

\begin{tabular}[c]{@{}l@{}}Wang \textit{et} \\ \textit{al.}~\cite{wang2020automated}\end{tabular}              & \begin{tabular}[c]{@{}l@{}}Pancreas and Tumor\\ Segmentation\end{tabular}                                                   & Horizontal FL                                                                                                    & \begin{tabular}[c]{@{}l@{}}NTUH Healthy Pancreas\\ CT (Taiwan), Nagoya-\\ Aichi Pancreatic Cancer\\ CT (Japan)\end{tabular} & \begin{tabular}[c]{@{}l@{}}First real-world FL implementation for pancreas\\ segmentation across international institutions\\ without data sharing.\end{tabular}                                                    \\ \hline

\begin{tabular}[c]{@{}l@{}}Shiri \textit{et} \\ \textit{al.}~\cite{shiri2023multi}\end{tabular}              & \begin{tabular}[c]{@{}l@{}}Head and Neck \\ Tumor Segmentation\\ in PET/CT\end{tabular}                                     & \begin{tabular}[c]{@{}l@{}}ClQu, ZeQu, FedAvg,\\ LoCo, RoAg, SeAg,\\ GDP-AQuCl\end{tabular}                      & HECKTOR dataset                                                                                                             & \begin{tabular}[c]{@{}l@{}}Demonstrate that pure transformer architectures\\ combined with FL achieve comparable performance\\ to centralized training for multi-institutional\\ PET/CT segmentation.\end{tabular} \\ \hline

\begin{tabular}[c]{@{}l@{}}Tolle \textit{et} \\ \textit{al.}~\cite{tolle2025real}\end{tabular}              & \begin{tabular}[c]{@{}l@{}}TAVI Landmark\\ Detection and \\ Segmentation\end{tabular}                                       & \begin{tabular}[c]{@{}l@{}}Semi-supervised\\ FedKD with SWIN-\\ UNETR\end{tabular}                               & \begin{tabular}[c]{@{}l@{}}Multi-center cardiac\\ CT, ImageCAS\end{tabular}                                                 & \begin{tabular}[c]{@{}l@{}}Combined FL with kD to leverage unlabeled data\\ for multi-task cardiac image analysis across\\ partially labeled datasets.\end{tabular}                                                 \\ \hline

\begin{tabular}[c]{@{}l@{}}Liu \textit{et} \\ \textit{al.}~\cite{liu2021feddg}\end{tabular}                & \begin{tabular}[c]{@{}l@{}}Federated Domain\\ Generalization for\\ Cross-site Segmentation\end{tabular}                     & FedAvg                                                                                                           & \begin{tabular}[c]{@{}l@{}}REFUGE, RIM-ONE,\\ Drishti-GS, NCI-ISBI\\ 2013, PROMISE12\end{tabular}                           & \begin{tabular}[c]{@{}l@{}}Enables FL models to generalize to unseen domains\\ via frequency-space interpolation and boundary-\\ aware episodic learning.\end{tabular}                                              \\ \hline

\begin{tabular}[c]{@{}l@{}}Jiang \textit{et} \\ \textit{al.}~\cite{jiang2023fair}\end{tabular}              & \begin{tabular}[c]{@{}l@{}}Medical Image\\ Segmentation with\\ Collaboration and\\ Performance Fairness\end{tabular}        & \begin{tabular}[c]{@{}l@{}}Contribution \\ Estimation-based\\ Aggregation\end{tabular}                           & REFUGE, PROMISE12                                                                                                           & \begin{tabular}[c]{@{}l@{}}Proposes dual-space client contribution estimation\\ (gradient direction and data error) to jointly\\ optimize fairness in FL medical segmentation.\end{tabular}                         \\ \hline

\begin{tabular}[c]{@{}l@{}}Galati \textit{et} \\ \textit{al.}~\cite{galati2024federated}\end{tabular}              & \begin{tabular}[c]{@{}l@{}}Multi-scanner Cardiac,\\  Multi-modality Skull\\Stripping, Multi-organ\\  vascular\end{tabular} & \begin{tabular}[c]{@{}l@{}}Federated multimodal\\ factory, Asynchronous\\  domain adaptation\end{tabular}        & \begin{tabular}[c]{@{}l@{}}M\&Ms(MRI), SynthStrip\\ (CT/PET), OASIS-3,\\ OCTA-500\end{tabular}                              & \begin{tabular}[c]{@{}l@{}}Proposes first FL framework supporting 4\\ modalities (MRI/CT/PET/OCTA) via disentangled \\ latent space.\end{tabular}                                                                   \\ \hline

\begin{tabular}[c]{@{}l@{}}Raggio \textit{et} \\ \textit{al.}~\cite{raggio2025fedsynthct}\end{tabular}           &   
\begin{tabular}[c]{@{}l@{}}Brain MRI-to-Synthetic \\CT Translation for Radio-\\ therapy Planning and PET \\ attenuation correction\end{tabular} 
& \begin{tabular}[c]{@{}l@{}}Cross-silo Horizontal\\ FL, 2D Residual U-Net, \\ FedAvg, FedProx\end{tabular}         
& \begin{tabular}[c]{@{}l@{}}Centres A(USA), B(IT),\\ C(DE), D(NL), E(NL, \\ SynthRAD2023)\end{tabular}                       
& \begin{tabular}[c]{@{}l@{}}Demonstrates a privacy-preserving FL approach for \\ multi-institutional brain MRI-to-CT synthesis\\ with robust generalization to unseen data.\end{tabular}\\   
\bottomrule
\end{tabular}
\label{table4}
\end{table*}

FL has emerged as a robust framework for collaborative segmentation of anatomical structures and pathological features in medical imaging~\cite{qayyum2022collaborative}. By facilitating the integration of distributed annotations, FL addresses key challenges, such as partial labeling and domain shifts. In multi-organ CT segmentation, Kim~\etal~\cite{kim2024federated} leveraged KD and uncertainty-aware aggregation to handle incomplete labels across multiple institutions. Their dual distillation strategy preserves organ-specific features by aligning local predictions with global feature representations. A lightweight multi-head U-Net architecture with shared encoders enables simultaneous segmentation of seven abdominal structures, mitigates catastrophic forgetting in non-IID settings while maintaining computational efficiency.

Building on this foundation, subsequent approaches, such as FUNAvg~\cite {tolle2024funavg}, introducing uncertainty-aware Bayesian aggregation into multi-organ segmentation. This method facilitates implicit knowledge transfer across institutions with incompatible labeling protocols, thus enhancing segmentation consistency and robustness in heterogeneous clinical environments. Galati~\etal~\cite{galati2024federated} propose a two‑stage federated segmentation framework for non‑IID multi‑center data, first learning a shared disentangled latent space via adversarial and contrastive objectives, then enabling asynchronous adaptation with limited local annotations and synthesized samples, achieving stable performance across diverse imaging tasks and uneven label distributions.

Liver-related segmentation tasks further illustrate the effectiveness of FL in harmonizing heterogeneous annotations across institutions. FedDUS~\cite{wang2024feddus} introduces a semi-supervised federated self-supervised learning (FSSL) framework, which incorporates dynamic client weighting and FAIR principles to standardize unlabeled CT data. To address domain shifts in liver tumor segmentation, federated adaptations of nnU-Net~\cite{luo2023influence} analyze inter-institutional distribution discrepancies and employ intensity normalization strategies to enhance cross-site consistency.

Hybrid cascaded models, such as Hybrid-ResUNet~\cite{hsiao2024precision}, combine 2D liver localization with 3D tumor delineation and employ transfer learning from related tumor datasets to improve model generalization for hepatocellular carcinoma segmentation. In pancreatic segmentation, Wang~\etal~\cite{wang2020automated} develop a federated framework designed to address domain shifts between healthy and pathological CT datasets collected from institutions in Taiwan and Japan. Their approach integrates neural architecture search with variational autoencoder regularization to align latent feature representations across institutions for robust multi-institutional segmentation.

Head and neck tumor segmentation frameworks incorporate vision transformers to capture long-range dependencies from PET/CT dual-modality data. These frameworks achieve performance comparable to centralized training through secure aggregation and DP mechanisms~\cite{shiri2023multi}. In cardiac CT analysis, semi-supervised FL strategies distill pseudo-labels derived from task-specific CNNs into transformer-based architectures, facilitating accurate landmark detection and calcification segmentation in sparsely annotated datasets~\cite{tolle2025real}.

FedDG~\cite{liu2021feddg} advances federated domain generalization for medical image segmentation by integrating frequency-space distribution interpolation with boundary-aware meta-learning. This framework supports collaborative training across decentralized, multi-hospital datasets while preserving data privacy by exchanging only the amplitude spectra of Fourier-transformed images and retaining the local phase spectra confidentially. The recombination of amplitude and phase components generates continuous style variations across domains, which enables the models to effectively generalize to unseen scanners and imaging protocols without exposing patient-sensitive information encoded within phase data.

FL also plays a critical role in enhancing fairness and robustness in medical image segmentation. Jiang~\etal~\cite{jiang2023fair} propose FedCE, which dynamically reweights client contributions according to gradient- and data-space metrics and reduces performance disparities in prostate MRI segmentation across institutions with varying imaging protocols. FL has also been extended to cross‑modality synthesis tasks. In brain MRI‑to‑CT conversion, Raggio~\etal~\cite{raggio2025fedsynthct} propose FedSynthCT‑Brain, a cross‑silo horizontal FL framework for multi‑institutional MRI‑to‑synthetic‑CT generation. By employing a residual U‑Net backbone and median‑voting fusion of axial, coronal, and sagittal predictions, their approach harmonizes heterogeneous imaging protocols without sharing raw data. Evaluation on both internal and external cohorts showed that FedSynthCT‑Brain maintains consistent image quality and anatomical accuracy across sites, demonstrating the feasibility of privacy‐preserving, federated cross‑modality synthesis in real‑world clinical settings.

These innovations collectively demonstrate the capability of FL to overcome annotation variability, data silos, and domain shifts in medical image segmentation, all while strictly adhering to privacy constraints.

\section{Future Perspectives}
\label{chap:5}
While significant progress has been made in applying FL to medical image reconstruction and analysis in recent years, several critical challenges remain unresolved. Future research should focus on enhancing model adaptability, strengthening privacy preservation, and improving computational efficiency across diverse medical imaging tasks. This section elaborates on these challenges and outlines potential avenues for future advancement.

\subsection{Special Characteristics of Medical Imaging}
Although existing methods have shown encouraging results, the majority still treat LDCT as a denoising problem, often neglecting the reconstruction process itself. This limits their compatibility to iterative unrolling-based CT reconstruction networks. In medical imaging, however, image formation is fundamentally based on reconstruction from measurement domain data guided by well-established physical models~\cite{xia2023physics}. Many existing frameworks fail to fully explore this measurement domain information, which constrains the optimization potential of reconstruction models. Iterative unrolling techniques, while grounded in physical modeling, typically require complex constraints and multi-step computations~\cite{xia2023regformer}, which substantially increase model complexity and computational burden.

Within the FL paradigm, these factors significantly amplify communication overhead and local computational demand, particularly on resource-limited edge devices such as mobile platforms and compact medical terminals. Such inefficiencies present substantial barriers to scalable training and real-world deployment in clinical settings.

Bridging this gap necessitates the development of FL-compatible iterative unrolling algorithms that balance high reconstruction fidelity with reduced computational and communication costs. Equally important is the effective integration of measurement domain data into FL-driven optimization, alongside the design of efficient and privacy-preserving client-server communication protocols. These advancements are essential for evolving current methodologies into fully-fledged reconstruction frameworks and ensuring their successful translation from theoretical innovation to clinical practice.

\subsection{Privacy Preservation Challenges}
Although FL is inherently designed to preserve data privacy by restricting raw data to local devices, it fundamentally relies on the assumption that all participants—both clients and servers—are fully trustworthy and free of malicious intent. In practical deployments, however, this assumption often fails to hold, exposing FL systems to a range of privacy threats. These risks are primarily manifested in the following aspects:

\begin{enumerate}
    \item While FL safeguards data privacy by transmitting only model updates, such as gradients or parameters, rather than raw data, it remains vulnerable to privacy leakage. Adversaries can exploit these updates using optimization-based reverse engineering techniques to infer client-specific inputs or latent features~\cite{zhu2019deep}. In particular, malicious entities may intercept gradients during transmission and reconstruct sensitive patient data, which poses a severe threat to confidentiality in FL-enabled medical systems.
    \item While current efforts mainly focus on data privacy protection~\cite{xue2021intellectual}, the intellectual property of model architectures remains under-protected and vulnerable to exploitation. This issue is particularly critical in cross-institutional collaborations and commercial deployments, where proprietary model designs and trained parameters can be subject to reverse engineering, unauthorized duplication, or malicious misuse.
    \item Pruned weight masks, which indicate retained and removed parameters, can inadvertently reveal client‑specific data distributions. As demonstrated by Yuan~\etal~\cite{yuan2022membership}, adversaries may infer membership of private samples by analyzing mask patterns. Furthermore, Chu~\etal~\cite{chu2025priprune} derive information‑theoretic bounds on leakage from pruned FL models and propose PriPrune, which demonstrates that naive pruning strategies are insufficient for privacy protection. These mask‑based attacks highlight the urgent need for privacy‑preserving pruning schemes—such as randomized or encrypted masks—in federated medical imaging scenarios.
\end{enumerate}

Although DP has been integrated into FL to safeguard gradient information, achieving an optimal trade-off between privacy preservation and model performance remains a persistent challenge~\cite{fukami2024dp}. Insufficient noise injection may fail to provide sufficient privacy protection, whereas excessive noise can significantly degrade model accuracy and hinder convergence during training. 

Alternatively, cryptographic techniques such as homomorphic encryption (HE)~\cite{acar2018survey} and secure multi-party computation (SMPC)~\cite{goldreich1998secure} provide strong privacy guarantees without compromising model performance, as they enable the aggregation of gradients or parameters without requiring decryption~\cite{xie2024efficiency}. However, these methods impose considerable computational and communication burdens~\cite{yang2025novel}, which can be prohibitive in resource-intensive tasks, such as medical image reconstruction, leading to prolonged training times and reduced scalability.

To address model privacy concerns, split learning has emerged as a promising alternative \cite{lin2024efficient}. Nevertheless, its inherently sequential training workflow offers lower efficiency than FL, limiting its scalability in large-scale distributed systems. Recent hybrid approaches, such as SplitFed~\cite {thapa2022splitfed}, aim to integrate split learning with FL to jointly preserve both data and model privacy while improving training efficiency via parallelism. However, these methods require the exchange of intermediate-layer features between clients and servers. Unlike high-level semantic tasks, medical image reconstruction is a low-level semantic task that typically avoids aggressive feature downsampling. As a result, the intermediate features are considerably larger in volume, which leads to communication overheads that far exceed those in conventional FL frameworks and potentially forms a critical bottleneck in real-world clinical deployment~\cite{yang2023dynamic}.

\subsection{Security Considerations}
\label{subsec:security}
DL algorithms, traditionally regarded as black-box models, have increasingly been identified as vulnerable to security threats such as backdoor attacks, necessitating robust safeguards in real-world deployments~\cite{yang2024inject}. These concerns are further exacerbated in FL frameworks due to their inherently decentralized training paradigm. The server’s inability to access raw client data limits the effectiveness of conventional security monitoring techniques and renders the detection and mitigation of adversarial behaviors significantly more challenging.

Moreover, the intrinsic heterogeneity among FL participants, including non-identical data distributions, varying computational resources, and inconsistent compliance with security protocols, undermines the ability to ensure uniformly benign behavior across clients. This architectural limitation exposes FL training to a wide range of security vulnerabilities, including data poisoning~\cite{gong2022backdoor} and Byzantine failures~\cite{so2020byzantine}, both of which can compromise the integrity and stability of the aggregated global model. As a result, reconciling the dual demands of strict data privacy and robust training security has emerged as a critical research priority in advancing FL systems.

\subsection{Communication Efficiency}
\label{subsec:communication}
Unlike conventional centralized training paradigms, FL operates in a distributed manner and requires careful management of communication costs throughout the training process~\cite{wu2024collaborative}. FL relies on frequent bidirectional communication between clients and a central server to exchange model updates~\cite{nguyen2021federated}. In the context of medical image analysis, characterized by high-dimensional data and models with large parameter counts, this leads to considerable communication overhead.

Compounding this issue, heterogeneity in client computational capabilities and unstable network conditions can cause asynchronous or delayed model uploads from certain participants. These disruptions hinder synchronized global model updates and may ultimately impede training efficiency~\cite{rong2022frcr}. To address these challenges, researchers have explored communication-efficient strategies, including model compression, sparse gradient transmission, and matrix factorization~\cite{shaofedufd}. These techniques are essential for scaling FL across large client networks in hospital systems and ensure that the deployment of FL remains feasible and efficient in real-world clinical environments.

\subsection{Scalability to Large Models}
Training parameter-intensive networks, such as 3D convolutional neural networks~\cite{yahiaoui2024federated} and vision transformers~\cite{alkhunaizi2024probing}, in an FL environment imposes substantial computational and communication demands on client devices. The size of model updates can be prohibitively large, which makes naive transmission over bandwidth-constrained networks impractical~\cite{wu2022communication,zhou2022communication}. To address these limitations, advanced strategies such as model compression, layer-wise training, and split learning are being actively explored to enhance the scalability and practicality of FL for large-scale models.
For instance, adaptive mutual KD techniques have demonstrated the ability to reduce communication costs by over 90\% with minimal impact on model accuracy~\cite{wu2022communication}. Federated model compression has emerged as a key solution for alleviating bandwidth and memory constraints~\cite{shah2021model,le2024survey}, thereby facilitating efficient deployment of high-capacity models in distributed healthcare environments. For example, pruning removes redundant parameters and then reduces the size of updates transmitted by each client. Stripelis~\etal~\cite{stripelis2022towards} propose FedSparsify, a method that prunes up to 95\% of model weights during federated brain age prediction without performance degradation on MRI data. Quantization compresses model weights into low‑bit formats, which cuts bandwidth consumption and accelerates local inference.  Gupta~\etal~\cite{gupta2024enhancing} applied int8 quantization via Quantization‑Aware Training to a 1D‑CNN for autism spectrum disorder classification on the ABIDE‑1 fMRI dataset, enabling edge deployment with minimal accuracy loss. Low‑Rank Adaptation attaches lightweight trainable adapters to a frozen backbone, which limits the size of transmitted updates to only a few megabytes. For instance, Zhu~\etal~\cite{zhu2024melo} develop MeLo, which adds only 0.17\% trainable parameters via LoRA to ViT models and achieves state‑of‑the‑art performance across multiple medical diagnosis tasks.
Knowledge distillation can be used to pre-train a compact student model prior to federated training, which reduces the resource requirements for initial deployment.  Kim~\etal~\cite{kim2024federated} used federated KD for multi‑organ CT segmentation on partially labeled datasets, which regularizes local training with global and organ‑specific teachers to improve accuracy with reduced student model size. Collectively, these techniques reduce network traffic and computational demands during federated rounds, and enable efficient training and real-time inference across heterogeneous hospital hardware.

Looking ahead, it will be essential to co‑optimize diffusion model compression with PACS bandwidth constraints, so that low‑precision or pruned diffusion priors can be streamed over DICOM networks without exceeding institutional data transfer limits. Real‑world edge deployments already demonstrate the potential of compressed models on resource‑constrained hardware platforms. Recent medical imaging case studies further validate this potential. Kromer~\etal~\cite{kromer2022medical} use lightweight CNNs (e.g., 6.25 MiB models) to achieve over 96\% diagnostic accuracy for COVID-19 detection on embedded NVIDIA Jetson modules, while optimizing energy-time tradeoffs to under 29W power consumption at 26.3 FPS inference rates. Blazeneo~\etal~\cite{an2022blazeneo} introduce BlazeNeo for real-time polyp segmentation and neoplasm detection,  and achieve over 155 FPS in INT8 precision on a Jetson AGX Xavier, while maintaining state‑of‑the‑art performance. Zhang~\etal~\cite{zhang2025adapting} adapt hierarchical vision foundation models for real‑time ultrasound image segmentation, achieving 77 FPS inference with TensorRT on a single A100 GPU and reporting a mean Dice score exceeding 90\% across six public and one in‑house dataset. Extending these case studies to federated, diffusion‑based reconstruction tasks will require end‑to‑end hardware–software co‑design frameworks that jointly satisfy compression, latency, and clinical workflow requirements.

\subsection{Post-Deployment Adaptation}
Deployed AI models in clinical settings must remain adaptive to evolving data distributions, driven by shifts in patient demographics, the introduction of new imaging devices, and updates to clinical protocols—factors that can lead to significant performance degradation due to distribution shifts. Federated Continual Learning provides a privacy-preserving framework for lifelong model updates across institutions~\cite{hamedi2025federated,koutsoubis2024future}. In this setting, each client periodically retrains on newly acquired local data and contributes model updates to a global model without disclosing sensitive patient information.

Crucially, recent studies highlight the importance of quantifying model uncertainty and detecting post-deployment distribution shifts, as exposure to unseen data can markedly compromise model reliability~\cite{koutsoubis2025privacy}. To promote robust and adaptive deployment, future research should prioritize integrating FL with domain adaptation and uncertainty estimation techniques, which enable models to autonomously recalibrate or retrain in response to novel clinical scenarios.

\subsection{Clinical Adoption Barriers}
The transition from validated federated models to clinical deployment encounters critical non-technical barriers. Hospital ethics committees (HECs) rigorously evaluate compliance with patient consent protocols, enforce data minimization principles, and assess the risk-benefit trade-offs involved in federated model training and deployment workflows. Simultaneously, data use agreements (DUAs) often require protracted inter-institutional negotiations, as they specify permissible data processing purposes, data retention periods, access rights, and ownership of jointly trained models, while ensuring compliance with regulations like the General Data Protection Regulation (GDPR)~\cite{mccall2018does} and the Health Insurance Portability and Accountability Act (HIPAA)~\cite{cohen2018hipaa}. These administrative processes often take 6 to 12 months, significantly delaying real-world validation.

Operational challenges further impede adoption. The allocation of liability for diagnostic errors remains ambiguous in FL, complicating institutional accountability and commitments. Cybersecurity certifications for hospital integration demand rigorous penetration testing of servers and encrypted communication channels~\cite{ahmed2022cybersecurity}. Seamless integration with clinical systems, such as Picture Archiving and Communication System (PACS), necessitates standardized interfaces to avoid workflow disruption. Clinician acceptance depends on transparent model interpretations for edge cases and proofable consistency in performance across clinical sites~\cite{metta2024towards}. Addressing these challenges through auditable federated architectures, standardized agreement templates, and clinician-centered interpretability tools is essential for transforming federated medical AI from a technical achievement to a clinically valuable asset.

\section{Conclusion}
FL has demonstrated significant promise for both the development and deployment of large-scale AI models for medical imaging. By transmitting only model updates, such as gradients or weight differences, rather than raw patient scans, FL enables collaborative training of AI models across various imaging modalities, including CT, MRI, and PET/CT. This approach not only safeguards patient privacy and adheres to regulatory constraints but also capitalizes on heterogeneous patient populations and imaging protocols that would be infeasible for any single institution to collect independently. Focusing on two core tasks in smart healthcare — medical image reconstruction and analysis, this review surveys recent FL-based methodologies, highlighting their strategies to handle non-IID data distributions, limited local data volumes, and the need for robust aggregation. At the same time, pronounced data heterogeneity and scarcity in medical imaging pose distinctive implementation challenges that often hinder model convergence and generalization. Accordingly, we systematically analyze these critical challenges and outline potential avenues to address them.

Sustaining the post-deployment performance of large models necessitates lightweight yet robust update mechanisms capable of incorporating new clinical cases and adapting to evolving imaging standards, all while minimizing downtime and mitigating the risk of data leakage. Future efforts should prioritize communication-efficient aggregation strategies such as sparse or quantized updates, model-parallel training schemes tailored for resource-limited sites, and federated continual learning protocols that detect distribution shifts and trigger safe model retraining. By addressing these engineering and governance challenges, FL can transition from proof-of-concept studies into a reliable framework for the responsible deployment and maintenance of next-generation AI systems across diverse healthcare settings.

\section*{Acknowledgment}
The authors declare that they have no known conflicts of interest in terms of competing 
financial interests or personal relationships that could have an influence or are relevant 
to the work reported in this paper.

\section*{Reference}
\bibliographystyle{IEEEtran}
\bibliography{refs}
\end{document}